\documentclass{book}    
\usepackage{piers}  
\begin{document}
\title{Releasable micro-waveplates}
\maketitle

\begin{authors}
 {\bf L. Grineviciute}$^{1}$, {\bf T. Tolenis}$^{1}$, {\bf M. Ryu}$^{2}$, {\bf T. Moein}$^{3}$, {\bf S.-H. Ng}$^{3}$, {\bf T. Katkus}$^{3}$, 
 {\bf Jovan Maksimovic$^{3}$,\\\protect  {\bf R. Drazdys}$^{1}$, {\bf J. Morikawa}$^{2}$, {\bf and S. Juodkazis}$^{3,4,5}$\\
\medskip
$^{1}$State Research Institute Center for Physical Sciences and Technology,
Savanoriu ave. 231, Vilnius LT-02300, Lithuania\\
$^{2}$Tokyo Institute of Technology, Meguro-ku, Tokyo 152-8550, Japan\\
$^{3}$Centre for Micro-Photonics, Swinburne University of Technology, John St., Hawthorn, VIC 3122, Australia\\
$^4$Tokyo Tech World Research Hub Initiative (WRHI), School of Materials and Chemical Technology, Tokyo Institute of Technology, 2-12-1, Ookayama, Meguro-ku, Tokyo 152-8550, Japan\\
$^{5}$Melbourne Centre for Nanofabrication, ANFF, 151 Wellington Road, Clayton, VIC 3168, Australia\\}
\end{authors}

\begin{paper}

\begin{piersabstract}
A simple procedure is demonstrated for fabrication of waveplates which can be released from substrate by laser cutting. Oblique angle deposition, chemical etching and laser inscription steps were used for the final lift-off and release of micro-waveplates in HCl solution.
\end{piersabstract}

\psection{Introduction}

Optical manipulation of materials using linear and angular (spin and orbital) momentum of light is utilised in laser tweezers, opto-mechanical manipulation of micro-objects~\cite{11}. Birefringence and its spatial variation are exploited to harness momentum of circularly polarised light. Spinning of birefringent particles - waveplates - is the most efficient when absorption is absent and birefringence causes the phase retardance of $\lambda/2$~\cite{2}. Deposition of linear and angular momentum of light by absorption is usually detrimental for a precise opto-mechanical manipulation due to heating.      
Here we show a solution how ultra-short pulse laser can be used to cut out an micro-optical element for subsequent opto-mechanical minipulation. For even smaller structures with cross section $< 10~\mu$m, focused ion beam (FIB, IonLine, Raith) milling can be used to cut through the waveplate region for a release of the micro-optical elements of required size (few-$\mu$m in cross section) and arbitrary shape. The waveplate elements were lifted off after laser inscription by a wet etch.

\begin{figure}[b]
\begin{center}
\includegraphics[width=16cm]{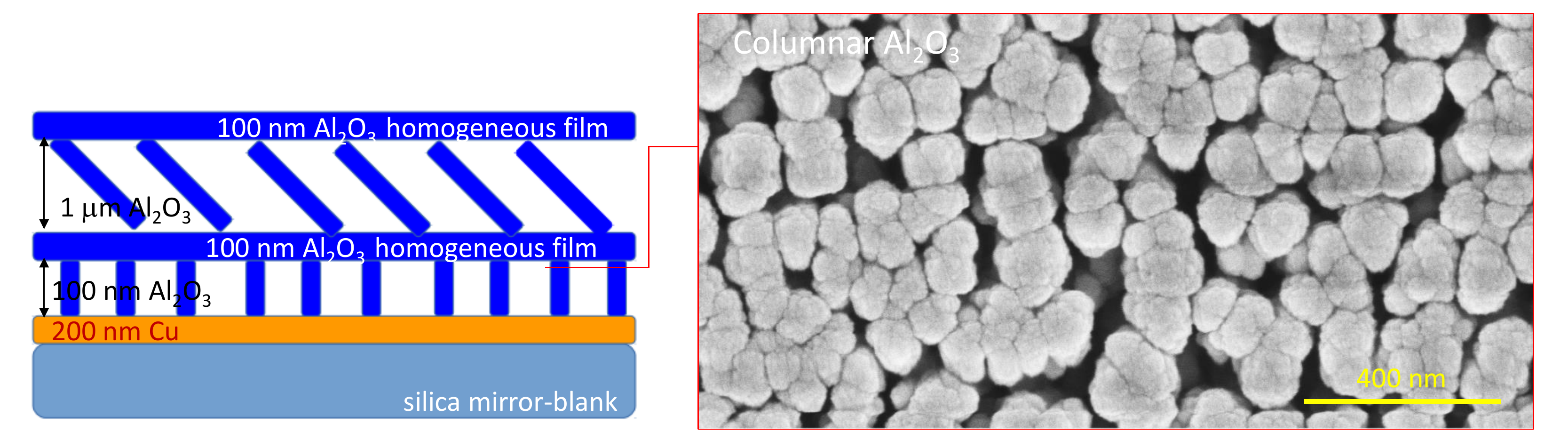}
\caption{A $\lambda/4$ plate for 515~nm wavelength (retardance of 128~nm) made by deposition of alumina layers of different porosity on a 200-nm-thick Cu film. SEM image (right) shows a top view image of the columnar alumina made to be permeable to acid etchant used for a lift off of the waveplate. }
\label{f-colu}
\end{center}
\end{figure}

\psection{Experimental: samples}

Waveplates were produced using an electron beam evaporation tool SIDRABE (Latvia). Fused silica mirror blanks were used as substrate. Porosity and nanostructure in the dielectric films were controlled by changing the deposition angle between substrate normal and vapor flux in high $P_0 = 1.5\times 10^{-3}$~Pa vacuum. 

Laser inscription of disk for the lift-off in 3M HCl solution were made by  $\lambda = 515$~nm wavelength, 230~fs duration pulses at 200~kHz repetition rate, using $\sim 0.3$~W average power (laser: Pharos, Light Conversion, Ltd.). The fluence was set twice as large as observable ablation threshold of Cu mirror which was $\sim 0.2$~J/cm$^2$/pulse. Focusing by an objective lens of $5^\times$ magnification and numerical aperture of $NA = 0.14$ was implemented; the diameter at the focus was $1.22\lambda/NA = 4.5~\mu$m. Inscription was carried out by five expanding rings (with a step of $\sim 2~\mu$m) between the passes at linear speed of 1~mm/s (with 250~pulses/mm density).

\psection{Results and discussion}

Columnar thin films (CTF), formed by oblique angle deposition technique, exhibited phase retardance due to the optical anisotropy~\cite{1}. The substrates were placed above the evaporation source and by manipulating it with dual stepper motors, variety of CTFs were realized (Fig.~\ref{f-colu}). Typical angle between the vapour flux and the substrate was 60-70$^\circ$. Fabricated multilayer structure consisted of copper film ($\sim$200-nm-thick) and four alumina layers: a columnar film ($\sim$100-nm), a dense supporting layer ($\sim$100-nm), an anisotropic columnar film ($\sim$1~$\mu$m waveplate) and a top dense sealing  layer ($\sim$100-nm same as the first supporting layer) as shown in Fig.~\ref{f-colu}. Alumina and copper films were formed by electron (e)-beam evaporation of pressed alumina powder and copper tablets, respectively. Time required for evaporation of CTF $\lambda/4$ plate was $\sim$3~hours. 

Retardance of the fabricated plates was measured with Berek compensator (Nichika Co., Japan No.10412; calibration constant for 550~nm $\log C = 3.856$) and is summarised in Fig.~\ref{f-reta} for two waveplates on Cu and birefringent layer deposited directly on silica blank. For the birefringenct CTF film on silica the retardance of $\Delta n\times d = 96$~nm was determined, which was close to the 128~nm required for $\lambda/4$ at $\lambda = 515$~nm. Retardance measured in reflection from the Cu film was only 60~nm (an optical double pass). This was, most probably, caused by comparatively thick non-birefingent layers of dense alumina on both sides of the CTF. The required thickness of the waveplates, dense sealing layers, and porous for under-etch is controlled only by deposition time and can be tuned to the required thicknesses.   

For the entire wavelpale lift-off, the substrate with a multilayer coating was immersed into HNO$_3$ water solution ($1:1$ vol.) to etch the copper layer and to release the wavelplate film from silica substrate. The first columnar alumina layer was permeable to etchant as can be seen from the top-view scanning electron microscopy (SEM) image (Fig.~\ref{f-colu}). It took $\sim 5$~min to release the entire waveplate. 

\begin{figure}[bt]
\begin{center}
\includegraphics[width=7.5cm]{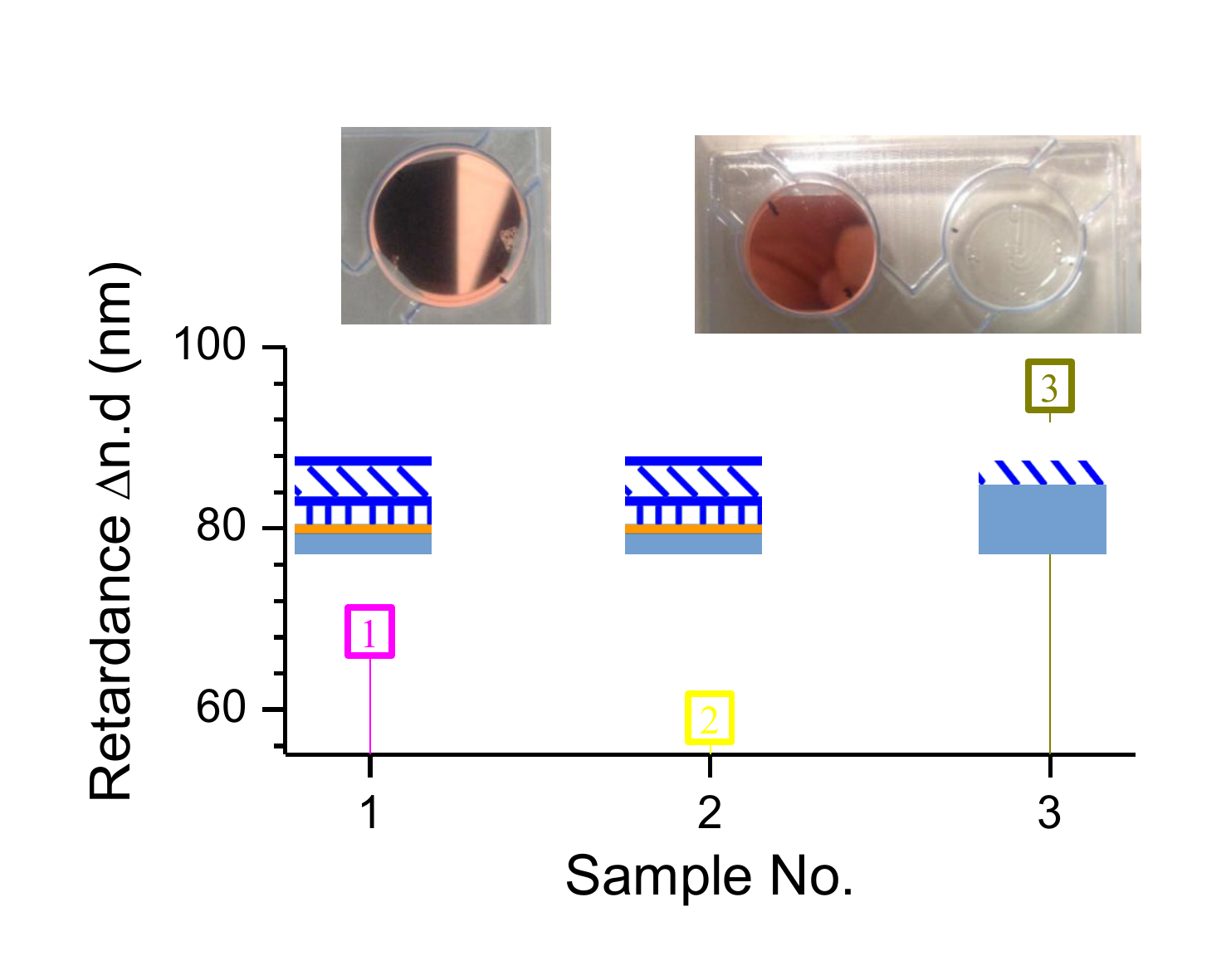}
\caption{Retardance measuerd by Bereck compensator (Nichika Co., Japan No.10412, calibration constant for 550~nm wavelength was $\log C = 3.856$) of different waveplates. Photo images and schematical structure are shown in the insets.}
\label{f-reta}
\end{center}
\end{figure}
\begin{figure}[tb]
\begin{center}
\includegraphics[width=15cm]{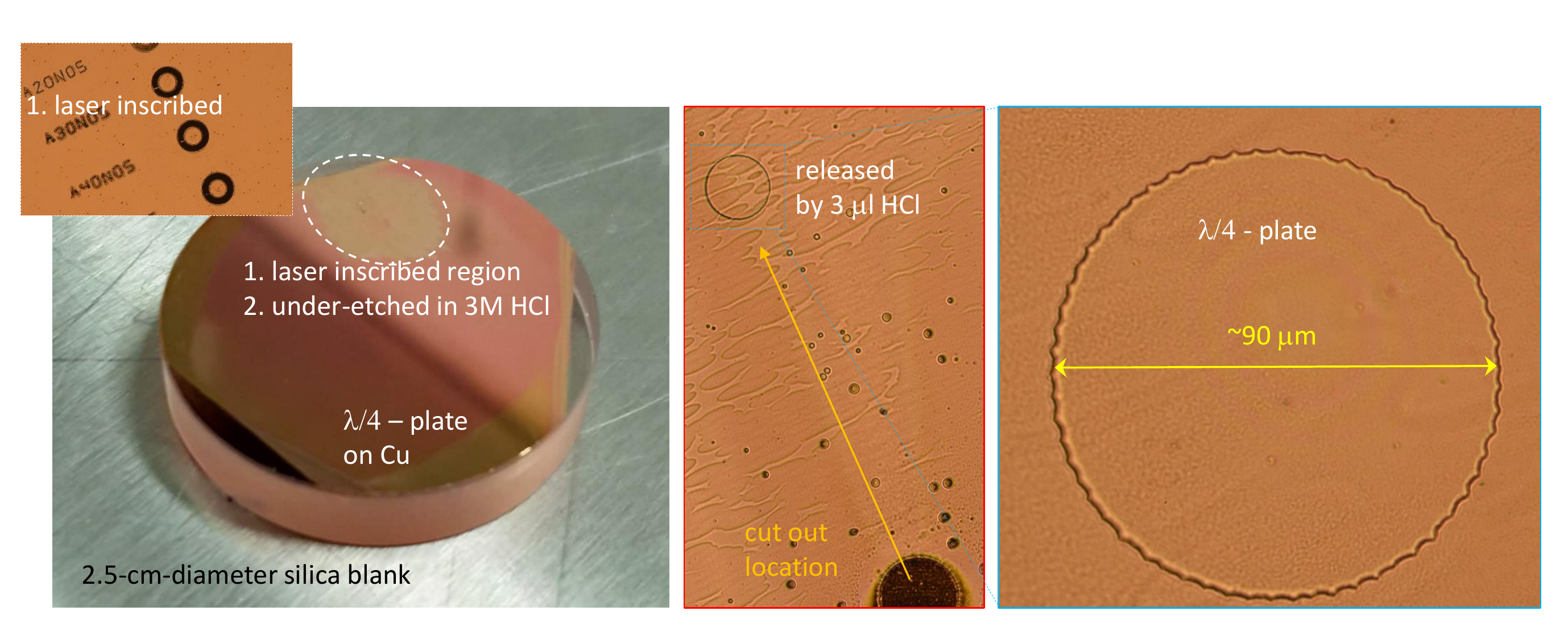}
\caption{A lift-off of a laser pre-cut waveplate by an under etch of Cu in HCl 3M solution. Laser cut was performed by 515~nm wavelength, 230~fs pulse duration pulses at 200~kHz repetition rate, scan speed of 1~mm/s under a numerical aperture $NA = 0.14$ focusing. The diameter of cutting trajectory was set to 100~$\mu$m diameter. After the release $\sim 90~\mu$m was obtained. The inset (top-left) show 40~$\mu$m discs after inscription. }
\label{f-under}
\end{center}
\end{figure}

For the release of micro-optical elements, the waveplate coating was inscribed with fs-laser pulses at the fluence which was twice larger for the observable ablation of Cu layer (Fig.~\ref{f-under}). Several concentric closely spaced circles were recorded to increase the surface area for permeation of the etching solution under the waveplate for removal of Cu. Down to 10-$\mu$m-diameter disks can be inscribed under used conditions in seconds. We used focusing with an objective lens of $NA = 0.14$ numerical aperture; the corresponding diffraction limit was $0.61\lambda/NA = 2.2~\mu$m (the inset in Fig.~\ref{f-under} shows inscribed regions of the 40-$\mu$m-diameter disks).     

For the test purpose, we used 3M HCl solution and only 3~$\mu$l droplet was placed over laser inscribed region. A real time waveplate release was observed under the microscope. High fidelity lift-off of all the waveplate disks was observed without side chipping of the edges (Fig.~\ref{f-under}). A well localised release of micro-waveplates was observed from the fabricated film. The sample can be reused for definition of other elements for subsequent lift-offs.

If smaller micro-optical structures should be fabricated for free manipulation by laser tweezers or laser beams on a liquid surface (interface), this can be achieved with focused ion beam (FIB) milling~\cite{fib}. The micro-objects released from Cu-film by the method shown in this study can be also used for nano-/micro-manipulation by strong light gradients created by patterns of plasmonic nanoparticles~\cite{XX}. 

A promising extension of this method is a possibility to use 3D laser polymerisation for fabrication of releasable 3D structures for opto-mechanical manipulation. The current method to achieve 3D-free objects by polymerisation requires an enclosure cage around the object, which is not attached to the surfaces of the substrate~\cite{move}. 
 
\psection{Conclusion}

The proposed procedure is simple and comparatively fast for fabrication of free-standing micro-optical elements used in polarisation manipulation at micro-scale. Such 3D-free and movable micro-structures (elements) are prospective for opto-mechanical manipulation using the linear and angular (spin and orbital) momentum of light.       
\ack
SJ acknowledges support via the Initiative of Excellence of the University of Bordeaux (IdEx Bordeaux) grant ANR-10-IDEX-03-02 and Workshop of Photonics, Ltd. for the process development grant.

\end{paper}

\end{document}